\documentclass[twocolumn,amssymb, nobibnotes, showpacs, superscriptaddress, aps, prd]{revtex4}
\usepackage{graphicx,amsmath,amssymb,color}

\begin{document}

\title{Nanotesla-level, shield-less, field-compensation-free, wave-mixing-enhanced body-temperature atomic magnetometry for biomagnetism} 

\author{Feng Zhou}
\affiliation{National Institute of Standards \& Technology, Gaithersburg, Maryland USA 20899}
\author{Eric Y. Zhu}
\affiliation{Department of Electrical and Computer Engineering, University of Toronto, Toronto, Ontario, M5S 3G4, Canada}
\author{Yvonne L. Li}
\affiliation{Dana-Farber Cancer Institute, Harvard Medical School, Boston, MA USA 02115}
\author{E. W. Hagley}
\affiliation{National Institute of Standards \& Technology, Gaithersburg, Maryland USA 20899}
\author{L. Deng}
\affiliation{National Institute of Standards \& Technology, Gaithersburg, Maryland USA 20899}
\email{lu.deng@nist.gov}

\date{\today}


\begin{abstract}
\noindent We report an optical inelastic-wave-mixing-enhanced atomic magnetometry technique that results in nT-level magnetic field detection at temperatures compatible with the human body without magnetic shielding, zero-field compensation, or high-frequency modulated phase-locking spectroscopy. Using Gaussian magnetic pulses that mimic the transient magnetic field produced by an action potential on a frog's nerve, we demonstrate more than 300,000-fold (550-fold) enhancement of magneto-optical rotation signal power spectral-density (power amplitude) over the conventional single-beam $\Lambda-$scheme atomic magnetometry method.  This new technique may bring possibilities for extremely sensitive magnetic field imaging of biological systems accessible via an optical fiber in clinical environments.
\end{abstract}

\maketitle

\vskip 5pt
\noindent
Human nervous system activities produce extremely weak transient magnetic fields at nano-Tesla (nT) to pico-Tesla (pT) levels during the exchange of information.  The ability to study these dynamic bio-magnetic impulses {\it in situ} under ambient conditions would therefore provide unique insight into real-time physiological processes. Starting with the pioneering works of Baule and McFee \cite{baule}, Cohen \cite{baule,cohen1,cohen2}, Wikswo et al. \cite{wikswo} and Gengerelli et al. \cite{gen}, this research field has significantly enriched our understanding of the role of magnetic fields in the human body. Indeed, the ability to measure magnetic fields generated by human organs and to track magnetic impulse evolution along various nervous systems have been key pursuits of the branch of bio-medical science referred to as bio-magnetism \cite{book1}.
One of the main, and perhaps most crucial, research thrusts in bio-magnetism is to improve our ability to detect extremely-weak magnetic fields under ambient conditions. This often requires new magnetometry technologies based on different physical and measurement principles. 
While mature technologies such as superconducting quantum interference {(SQUID)} magnetometry \cite{SQUIDS1,SQUIDS2,SQUIDS3} have been used in bio-magnetism, atomic magnetometry has been shown to be more sensitive \cite{sheng,Kominis}. Recent atomic magnetometric studies of human brain magnetic impulses \cite{BReview,brain}, the detection of cardiological activities of fetuses \cite{cardio1,cardio2,cardio3}, neurological disorders \cite{nerve1,kitching}, and the measurements of magnetic impulse generated in animal nerves by electric stimulations \cite{nerve2} demonstrate the great potential of atomic magnetometers.

\begin{figure}
  \centering
  \includegraphics[width=8 cm,angle=0]{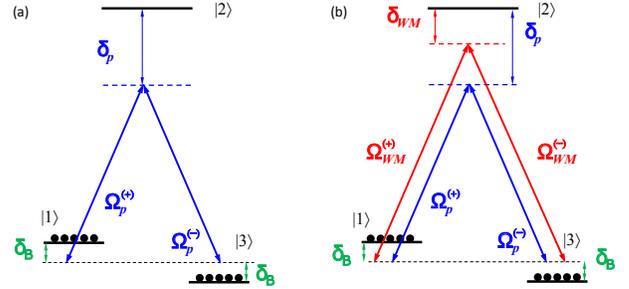}\\
\caption{{\bf Simplified atomic magnetometer schemes} (a) 
A generic single-beam $\Lambda-$scheme \cite{ref01}. The left- and right- circularly-polarized component of a linearly-polarized probe field forms two-photon transitions. The energy-symmetry restriction limits the growth of both components and the NMOR effect. (b) In the inelastic wave mixing scheme the energy from the WM field (red arrows) is transferred to the probe field through inelastic scattering process, removing the energy-symmetry based restriction and allowing a much larger NMOR effect. $\Omega_p^{(\pm)}$ ($\Omega_{WM}^{(\pm)}$) is the Rabi frequency of the probe (WM) field with corresponding detuning $\delta_p$ ($\delta_{WM}$). $\delta_B$ is the magnetic field induced Zeeman frequency shift. }
\end{figure}

\vskip 10pt
\noindent
Atomic magnetometry based on the magneto-optical rotation (MOR) effect measures the magnetic field dependent rotation of the polarization plane of a linearly-polarized probe light field traversing a magnetized atomic medium. Figure 1(a) shows the basic principle of this magnetometric method using a three-state atomic medium interacting with a probe laser field. This single-beam $\Lambda-$excitation configuration is widely used in many state-of-the-art atomic magnetometers \cite{book3}.  In this scheme the left- and right-circularly-polarized components of a linearly-polarized probe field jointly form two two-photon resonant channels, i.e., $|1\rangle\rightarrow|2\rangle\rightarrow|3\rangle$ and $|3\rangle\rightarrow|2\rangle\rightarrow|1\rangle$. Since the state $|1\rangle$ and state $|3\rangle$ are equally populated, these two competing two-photon transitions have the same probability by symmetry. However, because the linearly-polarized probe field is the only energy source, this symmetry property imposes a restriction on both two-photon transitions that neither component can change appreciably. This result in an energy-symmetry-based nonlinear propagation blockade that strongly limits the growth of both polarization components.  Correspondingly, the generation of the magnetic dichroism and Zeeman coherence at the probe frequency are strongly suppressed, resulting in a very weak probe field MOR effect. Consequently, all single-beam, three-state $\Lambda-$scheme based polarimetric atomic magnetometers, regardless of whether they are operating in a Raman-mode or an electromagnetically-induced transparency (EIT) mode \cite{book3,BReview,brain,cardio1,cardio2,cardio3,nerve1,kitching,nerve2,budker1,scully1,scully2,scully3,darkR3,darkR4,ref00,ref01}, have low optical polarimetric signal-to-noise ratios (SNR), {which must be overcome with} high medium density (temperature), high probe laser intensity, complex magnetic shielding \cite{shield}, zero-field compensation \cite{darkR4}, sophisticated phase-locking measurement protocols \cite{RF, RF2}, and long data acquisition times.
 
\vskip 10pt
\noindent Here, we demonstrate a novel nonlinear inelastic optical wave-mixing (WM) technique [Fig. 1(b)] for ultra sensitive atomic magnetometry \cite{zhu}. Physically, the WM field (red arrows) introduces a second excitation pathway that shares the same fully-populated intermediate Zeeman states with the probe field \cite{lutwm1,lureport}.  This results in a strong modification of the nonlinear dispersion and magnetic dichroism of the medium at the probe frequency. This shared-Zeeman-coherence at the probe frequency allows the energy to be transferred from the WM field to the probe field via inelastic WM and scattering, removing the energy-symmetry-based restriction on the growth of probe components and thereby breaking the self-limiting propagation blockade in single-beam $\Lambda-$schemes.  This results in a highly efficient MOR effect with unprecedented optical SNR enhancement \cite{exp} {not previously observed} in single-beam $\Lambda-$scheme atomic magnetometers. 

\vskip 10pt
\noindent In our experiments, we choose to compare the WM technique [Fig. 1(b)] side-by-side with the widely-studied single-beam $\Lambda-$scheme atomic magnetometer technique [Fig. 1(a)] first under our simple magnetic shield and then {\bf without} the use of shielding and zero-field compensation. No high frequency phase-locking detection electronics was used for all data reported in this work. In each set of data for comparison the {\it only} operational parameter change is the {absence} [Fig. 1(a)] or {presence} [Fig. 1(b)] of the WM field. This comparison is particularly instructive because the single-beam $\Lambda-$scheme atomic magnetometer has demonstrated {sensitivities comparable to that of SQUID devices}  {\it when state-of-the-art shieldings, field compensation, and high frequency phase-locking electronics are employed} \cite{sheng,BReview,Kominis}. 

\vskip 10pt
\noindent Figure 2 shows significant NMOR optical SNR enhancement effects obtained using the optical inelastic WM technique with a simple shield (the MOR effect often referred to as nonlinear magneto-optical rotation, NMOR, in literature and we will use this abbreviation hereafter).  As a comparison, we also show the performance of the conventional single-beam $\Lambda-$scheme atomic magnetometer under the same conditions. In these time-domain measurements we first measured the NMOR signal without the WM field (blue traces, right vertical scales).  This is exactly the single-beam $\Lambda-$scheme [Fig. 1(a)] widely-used in non-SERF-based atomic magnetometers which have demonstrated near $fT/\sqrt{Hz}$ level sensitivities at room temperature \cite{B1,B2} {\it when state-of-the-art magnetic shielding and phase-locking electronics are employed}. Typically, we choose the intensity of the probe field so that the NMOR signal of this conventional magnetometry scheme is at about the 2 mV level with a SNR $\approx$ 2 [Fig. 2(a), blue trace]. 
When a very weak WM field is introduced [see Fig. 1(b)], we routinely observe more than two orders of magnitude NMOR optical SNR enhancement [Fig. 2(a), red trace, left scale]. It is critical to note that the red trace with much higher signal amplitude has the {\it same} magnetic resonance line-shape as the blue trace, indicating that the WM technique {\it preserves} the magnetic resonance line-shape of the single-beam $\Lambda-$scheme. We emphasize that the only operational difference in {obtaining the two traces in} Fig. 2(a) was the {presence} or {absence} of the WM field.
When the probe field intensity is substantially reduced, no NMOR signal can be observed if the WM field is absent [Fig. 2(b), blue trace, right scale], even with extended acquisition times and data averaging. However, when a very weak WM field is turned on, a clear NMOR signal with an excellent SNR is observed [Fig. 2(b), red trace, left scale]. The {high-frequency} noise arises from oscilloscope electronic noise in the time-domain and can be substantially reduced by either using a spectrum analyzer or by averaging multiple scan traces with longer data acquisition times.

\begin{figure}
  \centering
  \includegraphics[width=8 cm,angle=0]{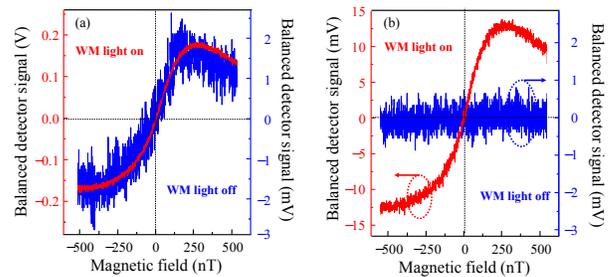}\\
  \caption{{\bf NMOR signal measured in the time domain at 311 K} (a): Red trace: a single scan of 25 ms with WM field. Blue trace: average of 128 scans in 3.2 seconds using the single-probe $\Lambda-$technique (i.e., without WM field) ($I_p$ = 284 $\mu$W/cm$^{2}$ with $\delta_p/2\pi = -$5 GHz and $I_{WM}$ = 80 $\mu$W/cm$^{2}$ with $\delta_{WM}/2\pi = -$2 GHz).  (b): NMOR signal with a substantially reduced probe field intensity.  Red/blue trace: with/without WM field ($I_p$ = 10 $\mu$W/cm$^{2}$ and $I_{WM}$ = 12 $\mu$W/cm$^{2}$). Fast noise on the blue trace is due to oscilloscope electronic noise. }
\end{figure}

\begin{figure}
  \centering
  \includegraphics[width=8 cm,angle=0]{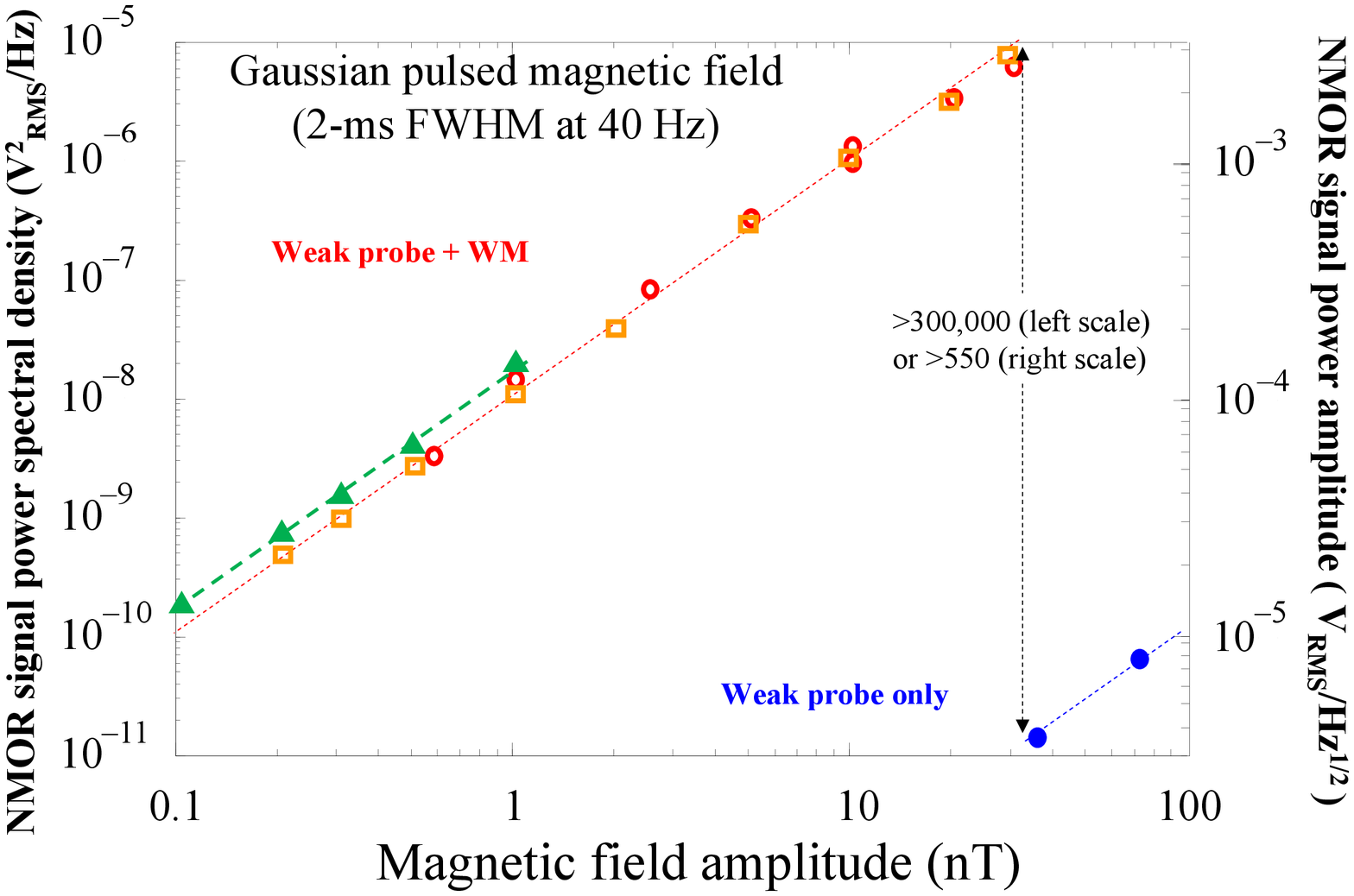}\\
  \caption{{\bf Peak NMOR signal power spectral density vs. magnetic field amplitude at 313 K with a simple shielding.} Blue dots: single-beam $\Lambda$ technique (no WM field) with $I_p= 1.5 mW/cm^2$. Red open-circles: optical WM technique using the same $I_p$ and $I_{WM}=60 \mu W/cm^2$. Orange squares: same parameters as in red circle but measured in a different day. Green triangles: same as orange squares, but with $I_{WM}=100 \mu W/cm^2$. All three dashed lines (blue, red, and green) have identical slopes, indicating that the WM technique preserves the magnetic field sensitivity but with $>\sqrt{300,000}\approx 550-$fold NMOR SNR enhancement. The Gaussian magnetic pulse with repetition rate of 40 Hz has 2-ms FWHM to mimic the magnetic impulses on a frog's nerve. }
\end{figure}

\vskip 10pt
\noindent
The detection of extremely weak magnetic impulses in biological systems is one of the primary bio-medical applications of atomic magnetometry. Neurons constantly fire electrical impulses while exchanging information, and this in turn generates extremely weak transient magnetic fields. Consequently, a technology with superior magnetic field impulse detection capabilities is advantageous in bio-magnetism research. A recent study \cite{nerve2} reporting detection of transient magnetic field impulses generated by an action potential on a frog nervous system demonstrated the great potential of atomic magnetometers.  The technique, however, rests on the conventional single-beam $\Lambda-$scheme atomic magnetometry, requiring high operational temperatures and a sophisticated magnetic field shielding environment. The optical WM atomic magnetometry reported here can substantially mitigate and reduce the dependency on these difficult operational conditions. In Fig. 3 we compare the NMOR optical signal power spectral densities of the optical WM technique and the single-beam $\Lambda-$scheme used in \cite{nerve2}.  Here, we operated at human body temperatures and we used a Gaussian magnetic field pulse train with a 2-ms full-width-at-half-maximum (FWHM) pulse width.  This format of magnetic pulses closely mimics the magnetic impulses in a frog's nerve generated by an action potential \cite{wikswo,nerve2}. The striking performance of the optical WM technique exhibits more than $300,000-$fold NMOR optical signal power spectral density enhancement (red circles, corresponding to about $\sqrt{300,000}\approx 550-$fold NMOR signal power amplitude SNR enhancement) in comparison with the conventional single-beam $\Lambda$ technique (blue dots). Experimentally, we have observed robust $>300,000-$fold NMOR optical signal power spectra density enhancements for long (100 ms) and short (200 $\mu$s) pulsed magnetic fields in square, sine and Gaussian wave modulation formats. We note that this unprecedented enhancement is achieved without even optical-pumping of the non-accessed hyper-fine states. We have observed experimentally a further enhancement factor of {more than} 2 with optical pumping (reaching $\approx 10^6-$fold in optical power spectra density enhancement, or $>$1000 fold signal amplitude enhancement), demonstrating the superior performance, large dynamic range, and robustness of the optical WM technique in transient magnetic impulse detection. We emphasize that the data from the optical WM technique (red squares and circles) in Fig. 3 exhibit {\bf perfect linearity}. Much more importantly, however, is the fact that the data from the optical WM technique exhibit the {\bf identical slope} as that from the single-beam $\Lambda-$technique (blue dots). This critically important feature has also been verified with sinusoidally-modulated magnetic fields.  As has been noted before {the} single-beam $\Lambda-$scheme has been shown to be able to produce $fT/\sqrt{Hz}$ level field sensitivity {\it when used with state-of-the-art shielding and phase-locking electronics}. This indicates that the optical WM scheme can reach the same detection sensitivity of the current state-of-the-art single-beam $\Lambda-$scheme magnetometers at 311 K under the same magnetic field shielding conditions, but with significant enhancement in the NMOR optical SNR. Using our simple magnetic shield and with a probe field of $I_p$ = 650 $\mu$W/cm$^2$ ($\delta_p=-2\pi\times$ 5 GHz) and a WM field of $I_{WM}$ = 60 $\mu$W/cm$^2$ ($\delta_{WM}=-2\pi\times$2 GHz) we have detected $19$ pT magnetic pulses with an NMOR signal power amplitude SNR $>$ 10 after averaging of only 64 scans \cite{notea}.  As a comparison, using the same shielding and electronics the single-beam $\Lambda-$scheme can only detect 5 nT pulsed magnetic fields with an NMOR signal signal amplitude SNR $\approx$ 8.  Extrapolating the slope of the optical WM data to the noise level of the system ($\approx 10^{-11}V^2/Hz$ with our spectrum analyzer), we expect to be able to detect 4 pT magnetic fields at an optical power amplitude SNR $\approx$1.
With state-of-the-art shielding and high-frequency phase-locking detection electronics, the body-temperature magnetic field detection limit shown here can be significantly improved.

\vskip 10pt
\noindent 
Current state-of-the-art NMOR-based atomic magnetometers have demonstrated near $fT/\sqrt{Hz}$ level magnetic field detection sensitivities in zero-field environments \cite{sheng,BReview,ref00,Kominis,B1,B2}. However, the requirements of complex and bulky magnetic field shielding \cite{shield}, cumbersome zero-field-compensating coils, and sophisticated high frequency modulated phase-locking electronics make them unsuitable for real-time high spatial resolution biomedical applications.  
Therefore, any new SNR enhancement technology that does not require sophisticated magnetic shielding will have great potential for bio-magnetism applications.  We note that without magnetic shielding the large background field already induces a large Zeeman frequency shift.  The non-zero-field NMOR signal due to a transient magnetic field arises from the extremely small magnetic dispersion and dichroism changes in the far wing of the corresponding two-photon resonance. 
The principle of this two-photon detuning sensing in the presence of the Earth's geomagnetic field is discussed in Fig. S4 in the Supplementary Information.

\begin{figure}
  \centering
  \includegraphics[width=8 cm,angle=0]{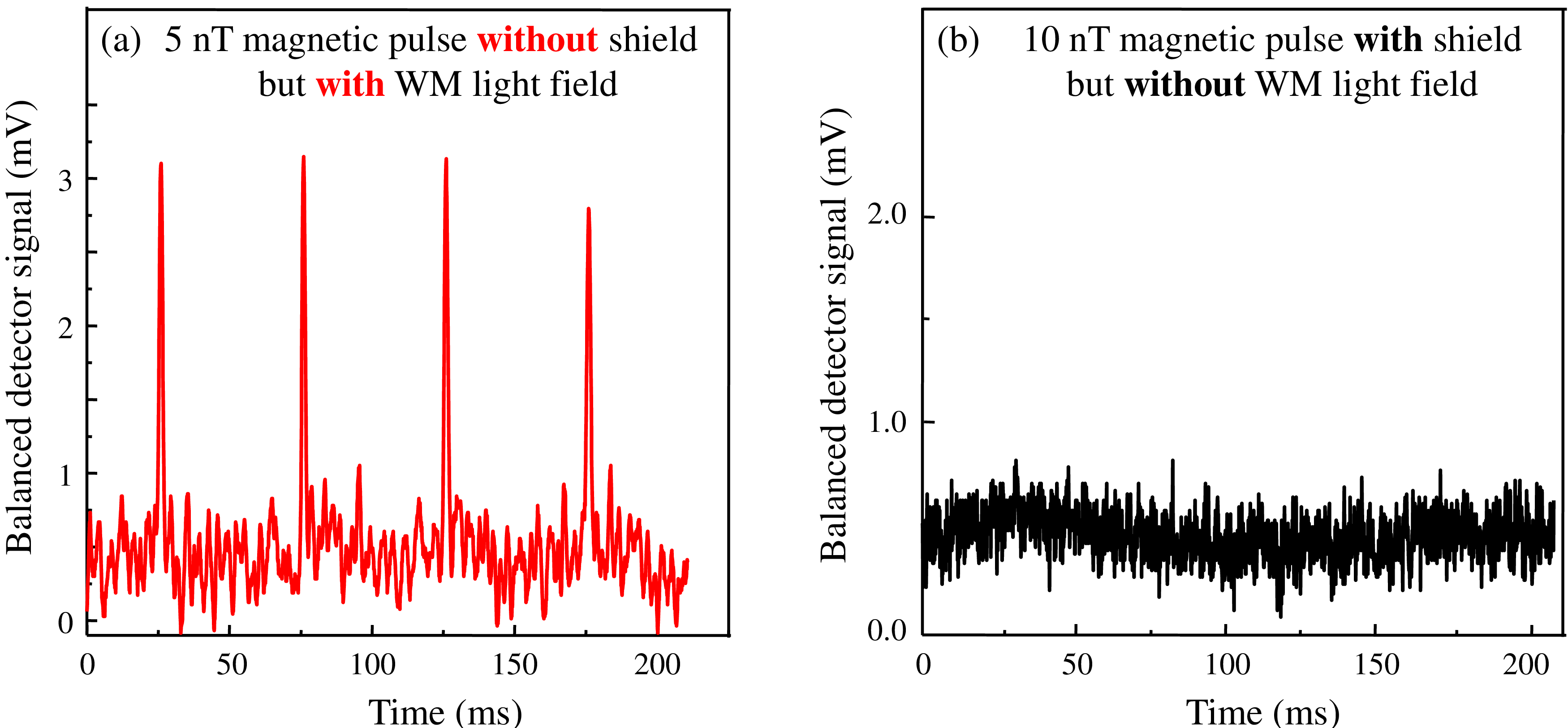}
  \caption{{\bf Pulsed magnetometry without magnetic field shielding or field compensation at 311 K.} (a):  5 nT pulsed magnetic field {\bf without} shield (0.8 nT has been observed with SNR = 1.5). (b): 10 nT pulsed magnetic field {\bf with} shield but {\bf without} WM field.  In all cases, no magnetic field compensation or phase-locking detection methods are employed. The intensities of the probe and WM lasers are 320 $\mu$W/cm$^2$ and 80 $\mu$W/cm$^2$, respectively. Data acquisition time (averaged over 64 scans) is about 14.4 second. }
\end{figure}
\vskip 10pt
\noindent Figure 4 demonstrates, in the time domain, weak pulsed magnetic field measurements {\bf without} any field shielding and zero-field compensation using our WM technique. This shield-less detection condition can realistically mimic the actual settings of typical biomedical laboratories and clinics. Here, the medium temperature is maintained at the human body temperature of 311 K (38$^o$C), and we input a train of square magnetic pulses{; Gaussian pulse trains of amplitudes varying from 0.5 nT to 5 nT, and 2 ms FWHM at 20 Hz were also used}. 
Without the WM field, no NMOR signal can be detected even with the use of a magnetic shielding or increased field amplitude of 10 nT [Fig. 4(b)].  However, when a weak WM field is turned on, the NMOR signal arising from the 5 nT magnetic field pulse is clearly visible {\bf without} shielding {or} field compensation [Fig. 4(a)]. 
Experimentally, { doubling the power of the probe allows us to measure a pulsed magnetic field of amplitude  0.8 nT with a SNR $\approx$ 1.5 in a shieldless environment simply by replacing the noisier oscilloscope with a spectrum analyzer (see Fig. 3).}   
We expect to reach $<$ 50 pT magnetic field detection {\bf without} magnetic shielding or zero field compensation at human body temperatures after further improvements to the optical system and by using high frequency phase-locking method. These estimates are consistent with the red dashed-line shown in Fig. 3 trending the behavior of the modulated WM magnetometry operation.

\vskip 10pt
\noindent 
The inelastic optical wave-mixing atomic magnetometry technology demonstrated here exhibits a superior NMOR optical SNR in weak magnetic field sensing applications. This technique may be used to study dynamic 3-dimensional magnetic field changes in physiological systems, as well as in disease and injury contexts such as neurological disorders and in the
recovery and regeneration phase of damaged nervous systems. Significant NMOR SNR enhancement also implies further reduction of the atomic magnetometer size to enable fiberization for endoscopic applications is possible. This would allow real-time {\it in situ} magnetic field mapping with substantially improved spatial resolution. We also note that because bio-magnetic abnormalities near tumors are anticipated given the distinct energetic molecular processes occurring in 
normal versus malignant cells, this highly sensitive technology may also provide a novel magnetic imaging platform for this new research direction. Finally, we note that since both the probe and WM fields operate in {continuous wave} modes the {measurement} is in a quasi-free-running mode waiting for the arrival of magnetic impulses. In essence, the technique acts as a free-running detector of transient magnetic fields with very high optical SNR, analogous to the LIGO gravitational wave detector which operationally outputs nothing until a gravitational (in our case a transient magnet pulse) passes by. 
It is for this reason the optical WM magnetometry technique may also have potential in the detection of extremely-weak magnetic fields generated by subatomic particles.


\vskip 10pt

\clearpage

\centerline{\bf Supplementary Materials}
\vskip 10pt
\noindent We first emphasize that the purpose of the present work is {\bf not} to demonstrate fT$/\sqrt{Hz}$ level sensitivities which requires a state-of-the-art zero-field shield/compensation environment with sophisticated high frequency modulated phase-locking electronics [S1,S2]. Our purpose is to demonstrate new physics which could potentially improve magnetometry without using magnetic shielding. If we can show the superior performance with a clearly-demonstrated performance trend when comparing with a well-studied state-of-the-art single-beam $\Lambda-$scheme-based atomic magnetometry technology under our experimental conditions, then it is reasonable to postulate that the large enhancement effect will occur when the state-of-the-art shield and electronics are employed. This is clearly shown by the perfect linearity and identical slopes in Fig. 3 of the text. In the following we provide more details on the experimental aspects of the new technique presented in this work.

\setcounter{figure}{0}
\makeatletter 
\renewcommand{\thefigure}{S\@arabic\c@figure}
\makeatother

\vskip 10pt
\noindent{\bf Experiment details}
\vskip 10pt
\noindent Figure S1 shows the energy levels of $^{87}$Rb used in our experiment and the experimental setup where identical NMOR effect measurement arms are designed for both the probe and WM fields. The symmetric two-arm setup allows the cross-reference of NMOR effects on both probe and WM fields.  For instance, by detecting the total energy change in each arm simultaneously, it is possible to conclude whether the underlying physics is a non-cyclic phase inelastic wave-mixing-based ground state Zeeman-coherence enhancement, or a double-$\Lambda$ type cyclic {four-wave-mixing} (FWM) process. 

\vskip 10pt
\noindent Experimentally, we use isotopically pure $^{87}$Rb atoms as the working medium. The atomic vapor is sealed in a cylindrical glass cell that is also filled with 933 Pascal of Neon buffer gas.  The uncoated cell is 5 cm in length and 2 cm in diameter.  The beam diameters for the probe and WM fields are 1.8 mm and 5 mm FWHM, respectively, and the typical operational temperature is 311-313 K (38-40 $^o$C). A solenoid previously calibrated using two-photon spectroscopy (by measuring field dependent two-photon resonance frequency shifts) is used to generate the axial magnetic field when the atomic vapor cell is coaxially located inside the solenoid cavity. We employed a simple home-made {magnetic shield} where two layers of $\mu-$metal form a cylindrically shaped housing with an inner spacing 30 cm in length and 12 cm in diameter. The shield limit is $\sim$500 nT (corresponding to $\delta_B \sim$ 3 kHz) in CW operation, as can be seen from Fig. 1a in the text. The temperature of the cell is maintained via convection and radiation of hot air at selected temperatures. The probe field couples the $5S_{1/2}$, $F = 2$ ground state manifold to the $5P_{1/2}$, $F^{'} = 1$ states with a detuning of $\delta_p/2\pi=-$5 GHz. The WM field couples the $5S_{1/2}$, $F = 2$ ground state manifold to the $5P_{1/2}$, $F^{'}$ = 1 states with a detuning of $\delta_{WM}/2\pi=-$2 GHz [S3]. We do not optically pump the non-accessed $F = 1$ ground state manifold and we measure the NMOR signal using a balanced detector in a typical polarimetry set up (detector bandwidth DC - 150 MHz). We have also studied all $D_1/D_2$ transition combinations experimentally. We must stress that the inelastic optical WM technique reported here is neither a pump-probe technique nor an electromagnetically-induced transparency technique [S3]. These techniques cannot produce large NMOR SNR enhancements because of the strong pump intensity required for transition saturation and transparency at the probe frequency.

\begin{figure}
  \centering
  \includegraphics[width=8.5 cm,angle=0]{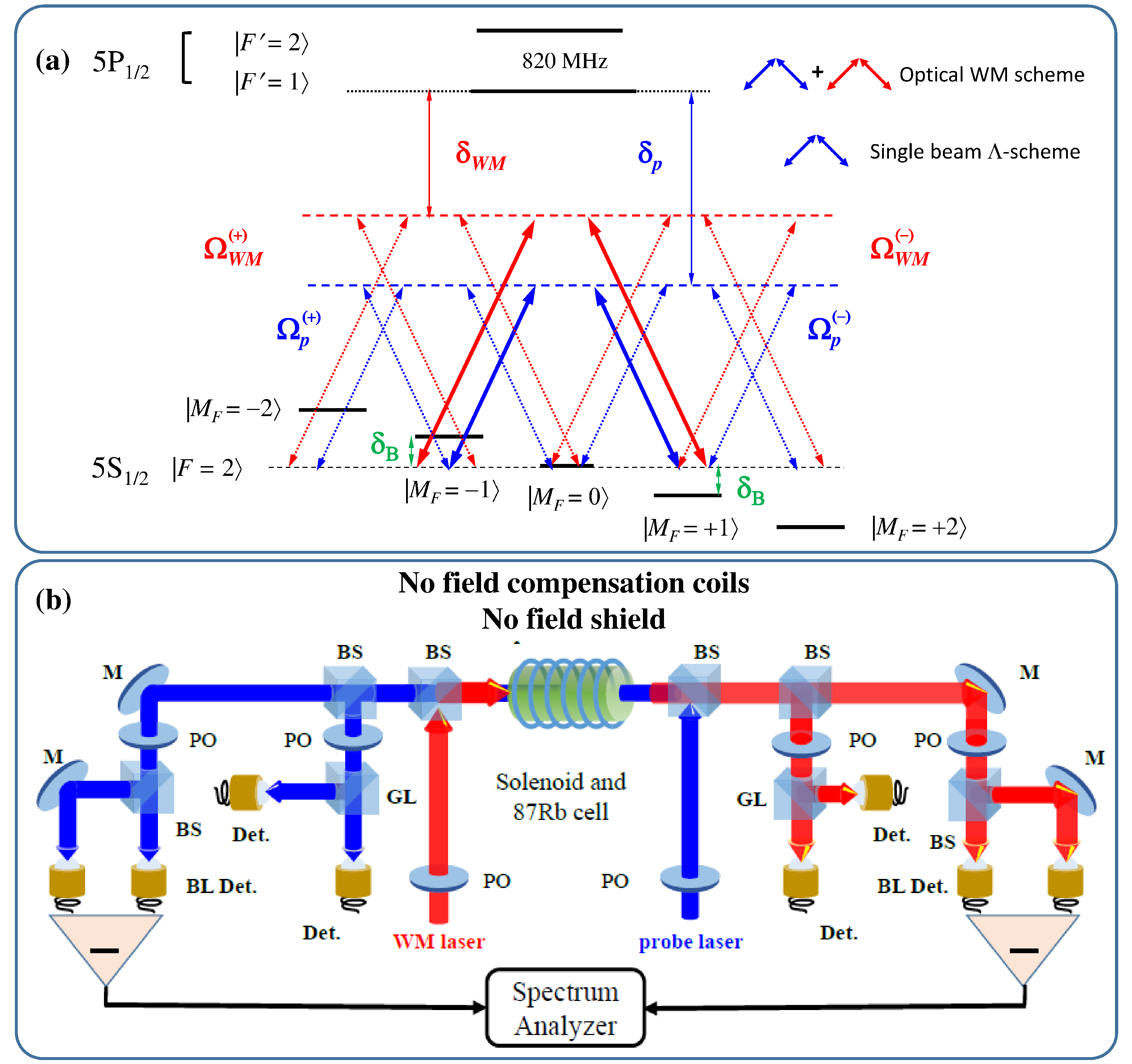}\\
  \caption{{\bf Atomic energy diagram and experimental setup.} (a) Simplified energy diagram for the most relevant transitions in $^{87}$Rb used in our experiment. The $F = 1$ hyperfine ground state manifold is not shown (not accessed). (b) Experimental arrangement where the symmetric detection setup for probe and WM field polarization rotations allows cross-reference of the NMOR effect observed in both arms. GP: Glenn prism, BD: balanced detector, D: detector, BS: beam splitter, DAQ/SA: data acquisition/spectrum analyzer.}
\end{figure}

\begin{figure}
  \centering
  \includegraphics[width=8.5 cm,angle=0]{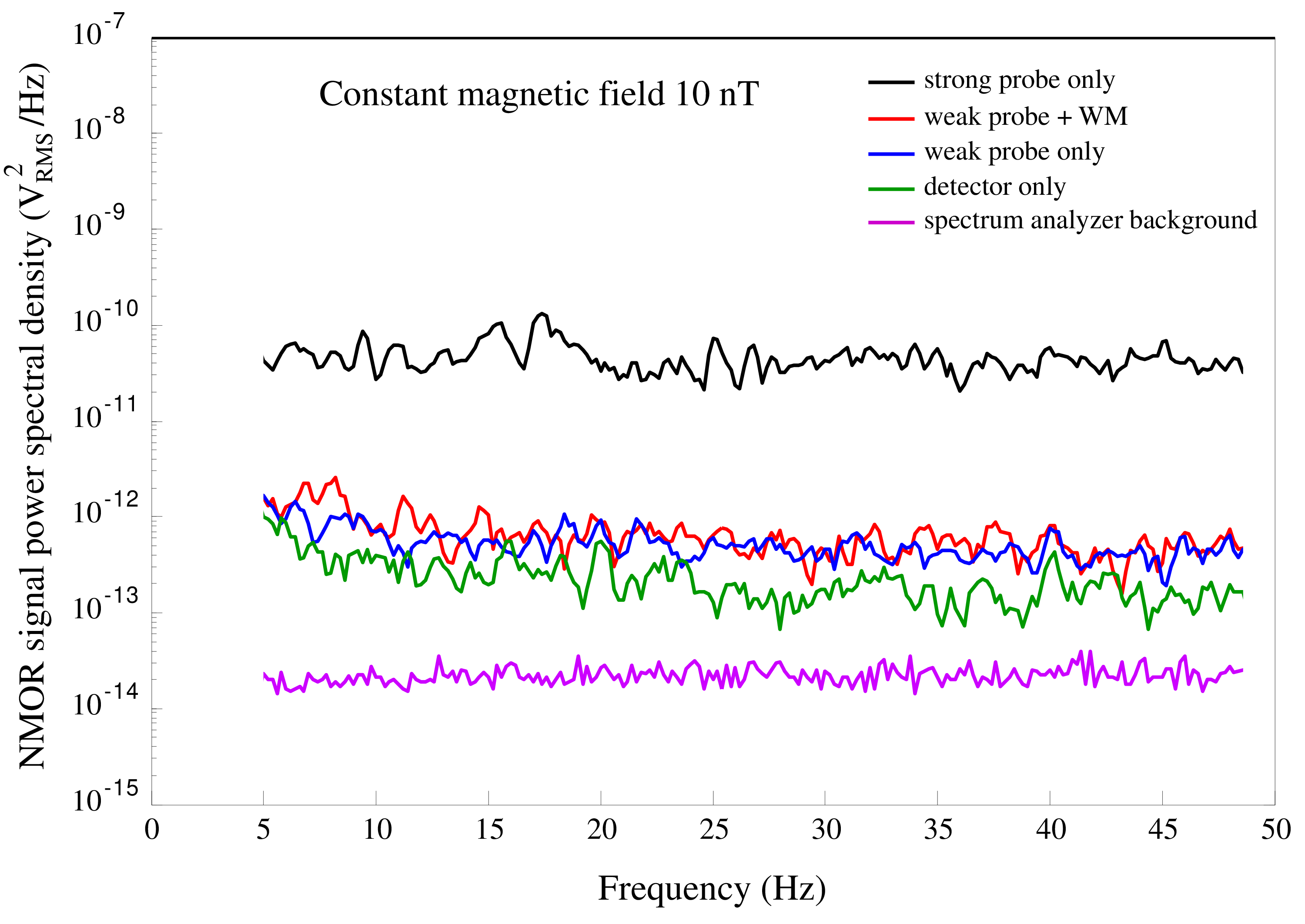}
\caption{{\bf NMOR optical signal power spectra densities at a constant magnetic field.} Blue trace: single-probe $\Lambda$ technique with a weak probe intensity of 17 $\mu$W/cm$^2$.  
Red trace: optical wave-mixing technique that yields a more-than 100-fold NMOR optical SNR enhancement in the time domain (similar to Fig. 1a). The probe and WM field intensities are 17 $\mu$W/cm$^2$ and 15 $\mu$W/cm$^2$, respectively. Black trace: single-probe $\Lambda$ technique using a high-power probe field (intensity 2.6 mW/cm$^2$) to generate a NMOR signal comparable to that of the WM technique.  Green trace: detector background. Purple trace: spectrum analyzer background. Data shown are the result of RMS averaging over 15 traces. The operational temperature is 313 K. The frequency region shown here is relevant to the pulsed modulation frequency.  We scanned DC-20 kHz range and made sure that there was no detectable high harmonic components.}
\end{figure}

\begin{figure}
  \centering
  \includegraphics[width=8.5 cm,angle=0]{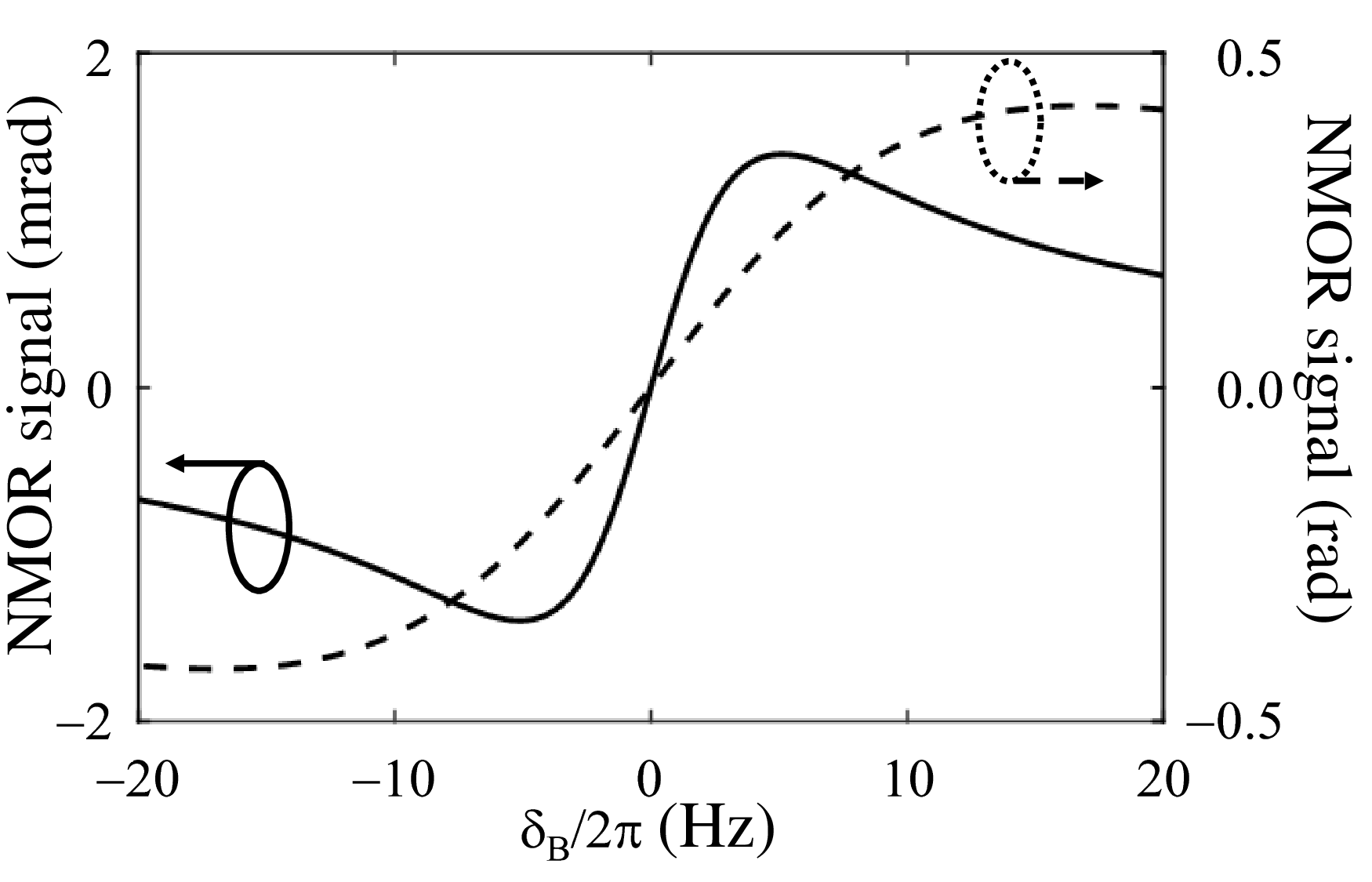}\\
  \caption{{\bf NMOR resonance line shape of the single-probe three-state $\Lambda-$scheme with different probe intensities.} Numerical calculations use the standard ellipsometry formalism. Solid line: weak probe: $\Omega_p^{(\pm)}/2\pi$ = 100 kHz. Dashed line: strong probe to generate a typical WM-scheme equivalent NMOR signal: $\Omega_p^{(\pm)}/2\pi$ = 1 MHz. The intrinsic Zeeman de-coherence rate is assumed to be 10 Hz.}
\end{figure}

\vskip 10pt
\noindent In a CW measurement, we typically set the magnetic field at a constant 10 nT and measure the signal power spectral density. A weak single beam is first used to generate the NMOR signal as in current state-of-the-art magnetometry.  The probe intensity is then reduced and a weak WM field is turned on.  We routinely observe 300 to 500-fold NMOR signal amplitude (i.e., optical {rotation}) enhancement in the time domain.  We then turned off the WM field and increased the probe intensity until the NMOR signal matched the amplitude generated with the weak probe and WM field.  In each of these cases, the NMOR optical noise power spectral density is measured. In Fig. S2 we demonstrate the superior low-noise performance by comparing the NMOR optical signal power spectral density for the three measurement schemes described using a spectrum analyzer.  Here, we measured the NMOR optical signal power spectra density at a constant magnetic field of 10 nT for three optical field excitation schemes: (1) a weak probe field only (i.e., the single-beam $\Lambda-$scheme in Fig. 1(a), blue trace); (2) a weak probe plus a weak WM field (the inelastic WM technique in Fig. 1(b), red trace); and (3) a strong probe field only (i.e., the single-beam $\Lambda-$scheme but with much higher intensity in order to produce an NMOR signal amplitude comparable to the inelastic WM technique, black trace).  First, it is amply clear that the noise of the optical inelastic WM technique (red trace) and the single-beam $\Lambda-$technique with the {\it same} weak probe field intensity (blue trace) are at the same level even though the inelastic WM technique produces an NMOR signal with more than two orders of magnitude larger amplitude.  This indicates that the WM technique does not introduce appreciable additional optical noise, a critically important technical advantage for applications in very low magnetic field detection. Secondly, in order for the single-beam $\Lambda$ technique to produce an NMOR signal amplitude comparable to that of the WM technique, the probe intensity must be substantially increased. This inevitably leads to a substantial increase in the optical noise level (black trace), along with power-broadening of the magnetic resonance that necessarily reduces the magnetic field detection sensitivity [S4].  Figure S2 demonstrates the superior noise-immune performance and great potential of the optical WM technique for bio-medical applications. We also note that the nearly 20 dB optical noise power spectral density difference exhibited in the NMOR signal spectra in Fig. S2 has also been observed over much broader frequency regions up to 20 kHz (the limit of our spectrum analyzer), demonstrating the broadband capability and robustness of the optical wave-mixing magnetometry technique. 

\vskip 10pt
\noindent In modulated magnetic field measurements, we first maximize the NMOR signal SNR enhancement in the time domain using an oscilloscope.  This yields CW mode results as shown in Fig. 2(a). We then fix the amplitude of the magnetic field, typically chosen in the range of 0.1 nT to 10 nT, and then modulate the magnetic field amplitude either sinusoidally or by a train of Gaussian pulses with a specified pulse FWHM (only the latter is reported in this work, see Fig. 3 in the text). The NMOR signal can be detected using either an oscilloscope (Fig. 2) or a spectrum analyzer (Fig. 3). The former is faster, but has substantial electronic noise whereas the latter has extremely low electronic noise but the data acquisition process is sinificantly slower. The spectrum analyzer has a frequency range of 0$-$25 kHz and the resolution bandwidth of 725 mHz is automatically selected by the analyzer for the sensitivity chosen in our experiment (using a flat-top filter). We first set the weak probe field intensity to obtain an NMOR signal with SNR $\approx$ 5 on the spectrum analyzer without the WM laser. For noise analysis we typically took four spectral traces as a constant magnetic field were collected in each batch (see Fig. S2).  We typically collected 15 batches and performed RMS averaging. We investigated a broad frequency range from DC to 20 kHz allowed by the spectrum analyzer. The WM field was then turned on and the intensity was adjusted to produce the maximum SNR enhancement. Typically, the intensity of the WM field is about a factor of 5 lower than the intensity of the probe. The noise spectra were similarly acquired and averaged. We then turned off the WM field and increased the intensity of the probe until the NMOR signal peak on the spectrum analyzer matched the peak height of the WM-enhanced NMOR signal previously measured. Typically, this requires an increase of the probe intensity by a factor of 50 or more. In the pulsed mode we made sure there was no detecable harmonic components.

\vskip 10pt 
\noindent We note that aside from the substantially increased optical noise as shown by the black trace in Fig. S2, a high probe intensity also leads to power-broadening of the magnetic resonance.  This inevitably reduces the field-detection sensitivity. Figure S3 shows Zeeman resonance line shapes of two different probe-field intensities where substantial magnetic resonance line broadening occurring with higher probe field intensities is clearly seen. The calculations are based on the single-beam, three-state $\Lambda-$scheme using the program given in [S5]. With parameters used in the present experiment, these calculations agree well with experimental observations and separate theoretical calculations based on nonlinear optics [S6].  

\vskip 10pt
\noindent 
In current state-of-the-art single-beam $\Lambda-$scheme atomic magnetometers, a high probe intensity with a frequency near the one-photon resonance is often required to improve the NMOR optical SNR. This is precisely the consequence of signal amplitude supression by the energy-symmetry blockade.  However, such high-power and small detuning conditions inevitably lead to power-broadening of magnetic resonance [S7] that degrades the detection sensitivity (as seen in Fig. S3). Reduced light field intensity results in reduced power broadening of the magnetic resonance and therefore the corresponding de-coherence rate of Zeeman states. It is well-known that the Zeeman de-coherence rate, in the absence of other relaxation mechanisms, determines the ultimate sensitivity of an atomic magnetometer [S4,S7]. The reduced magnetic resonance power broadening therefore is advantageous when the working medium has a very small Zeeman ground state de-coherence rate in the spin relaxation and destruction free regime. The optical WM technique demonstrated here, 
with its substantially larger one-photon detuning (typically 10$\times$ larger), low light intensity (typically 20$-$50$\times$ smaller), and superior NMOR optical SNR, therefore has great potential for very high resolution magnetic field sensing using atomic species with ground-state Zeeman de-coherence rates less than 1 Hz. Indeed, it is quite astonishing that the excitation rate used in our experiments are typically more than three orders of magnitude lower than that of usual single-beam $\Lambda$-scheme atomic magnetometry. 

\vskip 10pt
\noindent Finally, {even} in the presence of the Earth's magnetic field the WM-enhanced technique reported here 
{still has a significant SNR advantage}. The WM technique measures the extremely small two-photon detuning changes arising from minute medium magnetic dichroism changes in the far wing of two-photon resonances. Figure S4 describes this principle of two-photon detuning change sensing in the far wing of two-photon resonances in the presence of a large background Earth field induced Zeeman frequency shift.

\begin{figure}
  \centering
  \includegraphics[width=6 cm,angle=0]{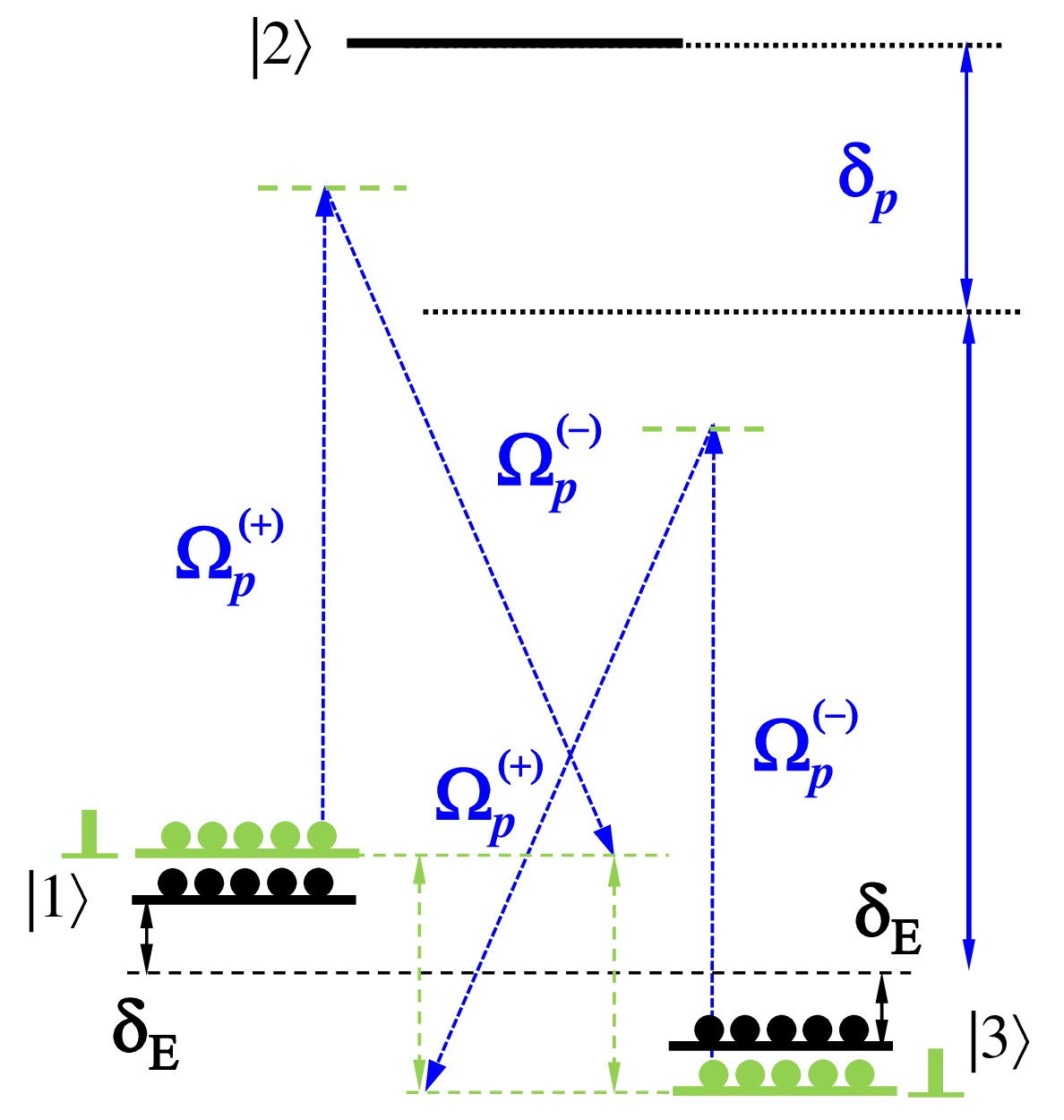}
  \caption{{\bf Laser coupling of pulsed magnetometry in the presence of a large background magnetic 
field (single beam configuration, WM field coupling can be similarly drawn)} 
  The background field introduces a large nearly static Zeeman shift {$\delta_E$}. The very weak pulsed magnetic field introduces a minute medium magnetic dichroism change (i.e., field dependent two-photon detuning change) which can be detected by the WM method due to the large SNR enhancement. 
  }
\end{figure}

\vskip 10pt
\noindent{\bf Further discussions}
\vskip 10pt
\noindent 
Further NMOR optical SNR enhancement by the optical WM technique can be obtained by optically pumping populations in the non-accessed ground state hyperfine manifold. Experimentally, we found that this population redistribution led to more than a 2-fold further increase in the NMOR signal amplitude, implying $>10^3-$fold NMOR optical signal amplitude SNR enhancement.  Choosing an atomic medium with a higher ambient vapor pressure and a larger dipole moment can further increase the overall NMOR sinal enhancement. For instance, by replacing $^{87}$Rb with $^{133}$Cs, similar NMOR signal SNR performance can be obtained in a cell only 2 mm in length. Therefore, a device with a mini-Cs-cell of 2 mm$^3$ in volume can provide NMOR optical signals with an excellent SNR at human body temperatures,  
opening the possibility of in-situ magnetic field sensing using an optical fiber for clinical endoscopy applications. 

\vskip 10pt
\noindent Phase-locking RF spectroscopy can further improve the NMOR optical SNR. Although such high-frequency probe-field modulation techniques can suppress random noise, or any other features that do not have a constant phase relation with respect to the probe, they are not well adapted to bio-medical applications. This is because there is no unique and yet passive synchronization reference in human nervous system that can serve as the primary reference clock for various nerve activities. Indeed, any imaging technology that requires precise and {\it active/external} probe-reference synchronization may not be suitable for studying transient biomagnetic signals and related dynamics. The optical WM technique demonstrated in this work can largely mitigate these difficulties since both the probe and WM fields operate in CW modes and the data acquisition is in a quasi-free-running mode waiting for the arrival of magnetic impulses. 

\vskip 25pt 
\noindent
{\bf References}
\vskip 10pt
\begin{itemize}
\item [S1.] I.K. Kominis, T.W. Kornack, J.C. Allred, and M.V. Romalis, Nature {\bf 422}, 596-599 (2003).
\item [S2.] D. Budker, D.J. Orlando, and V. Yashchukc, Am. J. Phys. {\bf 67}, 584-592 (1999).
\item [S3.] M.O. Scully, Phys. Rev. Lett. {\bf 67}, 1855-1858 (1991).
\item [S4.] J.F. Ward and A.V. Smith, Phys. Rev. Lett. {\bf 35}, 653-656 (1975). 
\item [S5.] M. Auzinsh, D. Budker, S. Rochester, {\it Optically Polarized Atoms}, Oxford University Press (2010).
\item [S6.] Y.R. Shen, {\it The Principles of Nonlinear Optics}, Chapters 13 - 16, John Wiley \& Sons, (1984).  
\item [S7.] For very small $\gamma_0$, the two-photon saturation effect of the strong probe required in current state-of-art technology begins to detrimentally broaden the Zeeman magnetic sub-levels, reducing the effective sensitivity.

\end{itemize}


\begin{thebibliography}{1}
\bibitem{notea} The minimum magnetic field strength {measured} in our experiment is limited primarily by the simple shield and by the time constraint of our slow spectrum analyzer.
\end{thebibliography}
\end{document}